\documentclass[twocolumn,showpacs,preprintnumbers,amsmath,amssymb,
 aps,
 prb,
 lengthcheck
]{revtex4-1}
\usepackage{amssymb}
\usepackage{graphicx}
\usepackage{dcolumn}
\usepackage{bm}
\usepackage{hyperref}
\usepackage{makeidx}

\usepackage[english]{babel} 
\usepackage[latin1]{inputenc}
\usepackage{color}
\usepackage{graphicx}
\usepackage{amsmath}

\usepackage{latexsym}
\usepackage{amsthm}

\makeindex
\def\>{\right\rangle}
\def\<{\left\langle}
\def\be{\begin{equation}}
\def\ee{\end{equation}}
\def\ba{\begin{array}{l}}
\def\ea{\end{array}}

\def\beq{\begin{eqnarray}}
\def\eeq{\end{eqnarray}} 

\begin{document}

\preprint{APS/123-QED}
 
\title{Tunneling between helical edge states through extended contacts}

\author{G. Dolcetto$^{1,2,3}$, S. Barbarino$^1$, D. Ferraro$^{1,2,3}$, N. Magnoli$^{1, 3}$, M. Sassetti$^{1, 2}$}
 \affiliation{$^1$ Dipartimento di Fisica, Universit\` a di Genova,Via Dodecaneso 33, 16146, Genova, Italy.\\
$^2$ CNR-SPIN, Via Dodecaneso 33, 16146, Genova, Italy.\\ 
$^3$ INFN, Via Dodecaneso 33, 16146, Genova, Italy.}
\date{\today}

\begin{abstract}
\noindent
We consider a quantum spin Hall system in a two-terminal setup, with an extended tunneling contact connecting upper and lower edges.
We analyze the effects of this geometry on the backscattering current as a function of voltage, temperature, and strength of the electron interactions.
We find that this configuration may be useful to confirm the helical nature of the edge states and to extract their propagation velocity. By comparing with the usual quantum point contact geometry, we observe that the power-law behaviors predicted for the backscattering current and the linear conductance are recovered for low enough energies, while different power-laws also emerge at higher energies. 

\end{abstract}

\maketitle
\section{Introduction}

Since the discovery of the Quantum Hall Effect (QHE) \cite{DasSarma}, the condensed matter community devoted great efforts in finding other topological states of matter in which fundamental physical properties are insensitive to smooth changes in material parameters and can be modified only through quantum phase transitions. In recent years, a new class of these peculiar systems have been experimentally observed: the topological insulators \cite{Hasan10, Qi11}. Their main characteristics are the presence of a gap in the bulk, analogous to the one of the ordinary band insulators, and gapless edge states protected by time-reversal symmetry. In two spatial dimensions they are the first realization of the Quantum Spin Hall Effect (QSHE), theoretically predicted in graphene with spin-orbit interaction \cite{Kane05a, Kane05b}, in strained semiconductors \cite{Bernevig06a} and in Mercury-Telluride quantum wells \cite{Bernevig06b, Konig07}. The edge states of the QSHE are helical \cite{Wu06}, namely their electrons have spin direction and momentum locked each other. In presence of intra-edge interactions they can be described in terms of a helical Luttinger liquid \cite{Wu06}. The experimental measurement of non-local transport in multi-terminal setups, according with the prediction of the Landauer-Buttiker theory \cite{Buttiker09}, represented an important test of the existence of helical edge states \cite{Roth09}.\\
The fast technical developments in this field will make shortly possible to realize interesting experimental geometries, like the Quantum Point Contact (QPC)  \cite{Chang03, Teo09, Czapkiewicz12} that already revealed extremely useful to extract information on the edge properties in the fractional QHE \cite{DePicciotto97, Chung03, Ferraro08, Ferraro10a, Ferraro10b, Carrega11}. Various theoretical proposals have investigated this geometry focusing on both the two-terminal \cite{Strom09} and four-terminal \cite{Hou09, Liu11, Schmidt11} setups. Possible interference experiments  \cite{Dolcini11, Virtanen11}, as well as quantum pumps \cite{Citro11}, involving two point contacts have also been considered.\\
The possibility offered by the Mercury-Telluride quantum wells to realize a QPC by means of electrostatic gates or, more realistically, by etching the sample in the desired shape makes possible to have a great control on the geometry and allows to study the evolution of the transport properties as a function of the constriction geometrical parameters. An analysis of the effects of extended contacts \cite{Aranzana05, Overbosch09, Chevallier10, Wang10} on the transport properties have been already addressed for the QHE showing deviations from the standard power-law behavior of the current as a function of the voltage at zero temperature.
Finite temperature effects were also considered for composite fractional QH systems \cite{Overbosch09}, demonstrating that extended contacts may provide information about the neutral mode propagation velocity along the edge, provided that it is very small with respect to the one of the charged mode.\\
In this paper we propose to investigate the extended contact geometry for the helical edge states of the QSHE by properly taking into account the role played by interactions. We will evaluate the backscattering current as a function of voltage and temperature. We will demonstrate that all the deviations with respect to the point-like case can be included in a modulating function. We will demonstrate that, at low enough temperatures, a peak appears in the differential conductance, which provides evidence of the helical nature of the edge states and gives information about the propagation velocity of the edge modes. At low energies the backscattering current and the linear conductance are described by the same power-law behaviors predicted for the QPC geometry. Even more interestingly, power-laws are recovered also at higher energies, but with different exponents.\\
The paper is divided as follows. In  Sec. \ref{model} we recall the main results of the helical Luttinger liquid description of edge states of a QSH system. In Sec. \ref{extended} we analyze the extended contact geometry introducing the modulating function both in the non-interacting and in the interacting case.   Sec. \ref{results} contains the main results on transport properties. Sec. \ref{conclusion} is devoted to the conclusions.

\section{Model} \label{model}

We consider a QSH insulator with one Kramers doublet of helical edge states in the two terminal configuration (see Fig. \ref{qlc}).
On the upper edge (1) one has right-moving spin up and left-moving spin down electrons, on the lower edge (2) the opposite.\newline
The corresponding free Hamiltonians are \cite{Hou09, Strom09} ($\hbar=1$)
\be
H_{1(2)}=-i v_{F} \int dx  \left(\psi^{\dagger}_{R, \uparrow (\downarrow)} \partial_{x}\psi_{R, \uparrow (\downarrow)}- \psi^{\dagger}_{L, \downarrow (\uparrow)} \partial_{x} \psi_{L, \downarrow (\uparrow)}\right)
\ee
where $\psi_{R, \uparrow}$ ($\psi_{L, \uparrow}$) annihilates right (left)-moving electron with spin up, and analogous for the spin down, and $v_{F}$ is the Fermi velocity, estimated\cite{Konig07, Goren96} about $5\cdot 10^5$ m/s. For sake of simplicity we assume infinite edges, even if a more realistic description based on finite length edges coupled to non-interacting leads can also be considered \cite{Liu11, Schmidt11}. This so called $g(x)$ model \cite{Maslov95, Safi95, Kleimann02} reveals crucial in order to recover the proper quantization of the conductance of one dimensional channels and leads to finite length corrections to physical quantities, that however are not crucial in the considered setup \cite{Liu11}.\newline
Concerning interactions, we consider terms which preserve time-reversal symmetry near the Fermi surface for a single Kramers doublet of helical edge states \cite{Moore06}. They are a subset of all possible contributions analyzed by the so called $g$-hology \cite{Giamarchi03, Miranda03} represented by the dispersive
\be\label{perp}
H_{d}= g_{2 \perp}\int dx\left(\psi^{\dagger}_{R, \uparrow} \psi_{R, \uparrow}\psi^{\dagger}_{L, \downarrow} \psi_{L, \downarrow}+\psi^{\dagger}_{L, \uparrow} \psi_{L, \uparrow}\psi^{\dagger}_{R, \downarrow} \psi_{R, \downarrow}\right)
\ee
and the forward scattering
\be\label{parallel}
H_{f}=\frac{g_{4 \parallel}}{2} \sum_{\alpha=R, L; \sigma=\uparrow, \downarrow} \int dx \psi^{\dagger}_{\alpha, \sigma} \psi_{\alpha, \sigma}\psi^{\dagger}_{\alpha, \sigma} \psi_{\alpha, \sigma}.
\ee
\newline
Note that possible Umklapp terms, which are important only at certain commensurate fillings \cite{Wu06}, are here neglected.\newline
The bosonized procedure of the Luttinger liquid allows to write the electronic field operator in the form \cite{Giamarchi03}
\be
\psi_{R/L,\sigma}(x)=\frac{\mathcal{F}_{R/L,\sigma}}{\sqrt{2 \pi a}}e^{\pm i k_{F}x} e^{-i\sqrt{2\pi} \varphi_{R/L,\sigma}(x)},
\ee
with $\varphi_{R/L,\sigma}(x)$ a bosonic field ($\sigma=\uparrow, \downarrow$), $\mathcal{F}_{R/L,\sigma}$ the Klein factor, necessary to give the proper commutation relation between electrons belonging to different edges, $a$ a finite length cut-off and $k_{F}$ the Fermi momentum. The bosonic field $\varphi_{R/L,\sigma}(x)$ is related to the electron density through $\rho_{R/L,\sigma}(x)=\mp\frac{1}{\sqrt{2\pi}}\partial_x\varphi_{R/L,\sigma}(x)$. According to the standard bosonization procedure \cite{Giamarchi03, Miranda03} the interaction terms in Eqs. (\ref{perp})-(\ref{parallel}) are quadratic in the electron density.\newline
Introducing the helical edge basis on the upper and lower edge \cite{Miranda03}
\be
\varphi_{1(2)}(x)=\frac{1}{\sqrt{2}}\left [\varphi_{L, \uparrow(\downarrow)}(x)-\varphi_{R, \downarrow(\uparrow)}(x)\right ],
\ee
with their canonical conjugates
\be
\theta_{1(2)}(x)=\frac{1}{\sqrt{2}}\left [\varphi_{L, \uparrow(\downarrow)}(x)+\varphi_{R, \downarrow(\uparrow)}(x)\right ],
\ee
the total Hamiltonian $H=H_1+H_2+H_{d}+H_{f}$ can be recast in the bosonized form \cite{Hou09, Strom09} 
\be
H=\frac{v}{2}\sum_{i=1, 2} \int dx \left[\frac{1}{K}\left(\partial_{x} \varphi_{i}\right)^{2}+K \left(\partial_{x}\theta_{i}\right)^{2}\right].
\ee
Here, $K=\sqrt{\frac{ 2\pi v_{F}+g_{4\parallel}-g_{2\perp}}{2\pi v_{F}+g_{4\parallel}+g_{2\perp}}}$ is the interaction parameter and $v= v_{F}\sqrt{\left(1+\frac{g_{4\parallel}}{2 \pi v_{F}}\right)^{2}-\left(\frac{g_{2\perp}}{ 2\pi v_{F}}\right)^{2}}$ the renormalized velocity.
For Coulomb repulsion $g_{4\parallel}=g_{2 \perp}$ and therefore $v=v_{F}/K$. In the following we will assume this condition, despite other possible interactions can be straightforwardly taken into account.

\section{Extended contact} \label{extended}

\begin{figure}[!ht]
\centering
\includegraphics[scale=0.28]{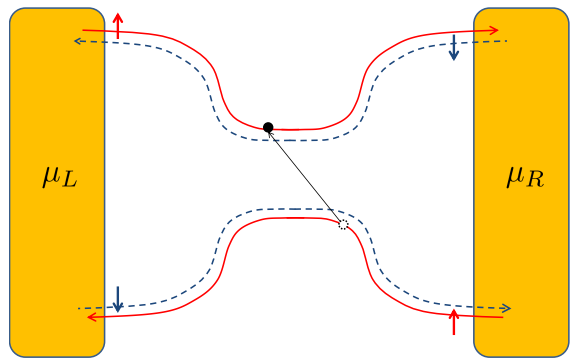}
\caption{(Color online) Extended contact geometry for a quantum spin Hall system with one Kramers doublet of helical edge states. The full (dashed) lines represent helical edge states carrying electrons with spin up (down). Right (left)-moving electrons are in equilibrium with the left (right) contact at chemical potential $\mu_{L}$ ($\mu_{R}$). The black arrow represents a possible spin-conserving electron tunneling event through the extended region.}\label{qlc}
\end{figure}

In presence of an external voltage $V$, right (left) -moving electrons feel a chemical potential $\mu_L$ ($\mu_R$), with $\mu_{L}-\mu_{R}=eV$. Spatial separation prevents electron tunneling between edges leading to conductance quantization \cite{Konig07} $G=\frac{e^2}{\pi}$. In order to study tunneling effects the system is pinched by means of a gate voltage \cite{Strom09} or, more realistically, by etching the sample \cite{Dolcini12} creating a tunneling region \cite{Chang03}.\newline
Previous theoretical works have studied this configuration \cite{Strom09, Teo09, Hou09, Liu11, Dolcini11}, both in two-terminal and in four-terminal setups, assuming a point-like tunneling.
In what follows, we will generalize this assumption, taking into account the possibility of tunneling events occurring in an extended region (see Fig. \ref{qlc}). Our aim is to investigate the effects induced by a long contact on the backscattering current.
The backscattering Hamiltonian connecting the two helical edge states is represented by 
\begin{equation}\label{H_tun}
H_{\mathrm{B}}=\int dx ~dy~  \left [ \sum_{\sigma=\uparrow, \downarrow} \Lambda_{x,y}\psi^{\dagger}_{R,\sigma}(x)\psi_{L,\sigma}(y)\right]+h.c.,
\end{equation}
with $\Lambda_{x,y}$ the tunneling amplitude in which a left-moving electron is destroyed in $y$ on one edge and recreated as a right-moving electron in $x$ on the other one. A reasonable choice for $\Lambda_{x,y}$ is to assume a separable form \cite{Chevallier10}
\begin{equation}
\Lambda_{x,y}=\Lambda_0f_l\left (|x+y|\right )  f_c\left(|x-y|\right).
\label{separabile}
\end{equation}
The function $f_{l}$, indicated as lateral contribution, specifies the average location of the tunneling events \cite{Aranzana05, Overbosch09, Chevallier10} while $f_{c}$, dubbed crossed, allows to take into account non perfectly vertical events \cite{Chevallier10}. This assumption is reasonable for smooth tunneling junctions.
Both functions are maximal around zero and decrease by increasing their arguments. With this requirement, the longer the tunneling path the smaller the corresponding local amplitude.\\
Note that Eq. (\ref{H_tun}) describes spin-conserving tunneling processes only, since spin-flipping tunneling terms give no contribution in our two-terminal setup \cite{Liu11, Dolcini11}. Furthermore we neglect tunneling of either charged ($\sim \cos\left [\frac{1}{\sqrt{\pi}}\left (\varphi_1+\varphi_2\right )\right ]$) or spinful ($\sim \cos\left [\frac{1}{\sqrt{\pi}}\left (\theta_1-\theta_2\right )\right ]$) particle pairs, although for strong enough electron interactions they could compete with single-particle tunneling processes ($\sim \cos\left [\frac{1}{2\sqrt{\pi}}\left (\varphi_1+\varphi_2\right )\right ]\cos\left [\frac{1}{2\sqrt{\pi}}\left (\theta_1-\theta_2\right )\right ]$) \cite{Teo09, Schmidt11}. Note that all these processes are irrelevant \cite{Teo09}, in the RG sense, for $0.5 < K < 2$. We limit our analysis to repulsive interaction $0.5 < K < 1$, and we treat the tunneling current as a small perturbation.\\
The tunneling Hamiltonian in Eq. \eqref{H_tun} induces no net charge transfer between the two edges, but leads to a net spin tunneling current.
The corresponding spin current operator is  
\be
I_{\mathrm{S}}=-\frac{i}{2} \sum_{\sigma=\uparrow, \downarrow}\int dx~dy~\Lambda_{x, y}\psi^{\dagger}_{R,\sigma}(x)\psi_{L,\sigma}(y)+h.c.,
\label{current_operator}
\ee
according to the requirement of absence of spin flipping and multiple-particles contributions. In linear response approximation in the tunneling hamiltonian, the stationary expectation value of the spin current in Eq. (\ref{current_operator}) can be written in terms of the tunneling rates ${\bf{\Gamma}}_{L,\sigma\to R, \sigma}$ and ${\bf{\Gamma}}_{R,\sigma\to L,\sigma}$ as  
\be
\label{I_S}
\langle I_{\mathrm{S}} \rangle=\frac{1}{2}\sum_{\sigma=\uparrow,\downarrow}\left [ {\bf{\Gamma}}_{L,\sigma\to R,\sigma}-{\bf{\Gamma}}_{R,\sigma\to L,\sigma}\right ] .
\ee
Note that the functional dependence of rates and other physical quantities from bias and temperature is understood for notational convenience.
 
One can easily realize that this spin tunneling current is responsible for a reduction of the net current flowing from one lead to the other \cite{Strom09, Liu11}, i.e. $\langle I_{\mathrm{}} \rangle=\frac{e^2}{\pi}V-\langle I_{\mathrm{BS}} \rangle$, with $\langle I_{\mathrm{BS}} \rangle$ the backscattering current, related to $\langle I_{\mathrm{S}} \rangle$ by
\be
\langle I_{\mathrm{BS}}\rangle=2e\langle I_{\mathrm{S}}\rangle.
\ee
We can thus measure the spin tunneling current by measuring the ordinary backscattering current \cite{Strom09}.\\
By taking into account the spin independence of the tunneling rates and by considering the detailed balance relation ${\bf{\Gamma}}_{R,\sigma\to L,\sigma}=e^{-\beta eV}{\bf{\Gamma}}_{L, \sigma\to R,\sigma}$, ($\beta=1/k_{B}T$ the inverse temperature) one has
\be
\langle I_{\mathrm{BS}}\rangle= 2e\left (1-e^{-\beta eV}\right ) {\bf{\Gamma}}_{L,\uparrow\to R,\uparrow}.
\label{current_detailed}
\ee
According to Eq. (\ref{current_detailed}), we can consider only the tunneling rate ${\bf{\Gamma}}\equiv {\bf{\Gamma}}_{L,\uparrow\to R,\uparrow}$ given by
\beq
{\bf{\Gamma}}&=&\int dx~dy~dx'~dy'~\Lambda_{x,y}\Lambda_{x',y'}^{*}\nonumber \\
&\times&\int dt~e^{i eVt}G^{>}_{L}(y'-y,t)G^{<}_{R}(x'-x,t), 
\label{qlcrrate}
\eeq
with
\beq
G_{R/L}^>(x,t)&=&\frac{e^{\mp ik_Fx}}{2\pi a}e^{\mathcal{W}_{R/L}(x,t)}\\
G_{R/L}^<(x,t)&=&\frac{e^{\pm ik_Fx}}{2\pi a}e^{\mathcal{W}_{R/L}(x,t)}
\eeq
the greater and lesser electron Green's functions associated to the right $(R)$ and left $(L)$ movers. The corresponding bosonic Green's functions  are
\begin{eqnarray}
\mathcal{W}_{R/L}(x,t)&=&2\pi\langle \varphi_{R/L, \sigma}(x, t) \varphi_{R/L, \sigma}(0,0)\rangle \nonumber \\
& &-2\pi \langle \varphi_{R/L, \sigma}(0, 0) \varphi_{R/L, \sigma}(0,0)\rangle.
\end{eqnarray}
They do not depend on spin and can be written in terms of the chiral ones $\mathcal{W}_{\pm}(x,t)$
\begin{eqnarray}
\mathcal{W}_R(x,t)&=&c^{(+)}_K \mathcal{W}_+(x,t)+c^{(-)}_K\mathcal{W}_-(x,t)\\
\mathcal{W}_L(x,t)&=&c^{(-)}_K\mathcal{W}_+(x,t)+c^{(+)}_K\mathcal{W}_-(x,t),
\end{eqnarray}
with 
\begin{equation}
\mathcal{W}_{\pm}(x,t)=\mathcal{W}\left(t\mp \frac{x}{v} \right)
\end{equation}
and 
\begin{equation}
\mathcal{W}(t)=\ln\left [\frac{\left |\Gamma\left (1+\frac{1}{\beta\omega_{c}}-i\frac{t}{\beta}\right )\right |^2}{\Gamma^2\left (1+\frac{1}{\beta\omega_{c}}\right )\left (1+i\omega_{c} t\right )}\right ].
\label{W_exact}
\end{equation}
Here, $\Gamma(x)$ is the Euler Gamma function,  $c^{(\pm)}_K=\frac{1}{4}\left (\sqrt{K}\pm\frac{1}{\sqrt{K}}\right )^2$ are the interaction dependent tunneling coefficients  and $\omega_{c} =v/a$ the energy bandwidth. By replacing the above expressions into  Eq. \eqref{qlcrrate} one obtains
\begin{widetext}
\be
{\bf{\Gamma}}_{K}=\int dx~dy~dx'~dy'~ \frac{\Lambda_{x,y}\Lambda_{x',y'}^*}{(2\pi a)^2}e^{ik_F(y'-y+x'-x)}\int dt~e^{ieV t} e^{c^{(+)}_K\mathcal{W}(t-\frac{x'-x}{v}) +c^{(-)}_K\mathcal{W}(t+\frac{x'-x}{v}) +c^{(-)}_K\mathcal{W}(t-\frac{y'-y}{v}) +c^{(+)}_K\mathcal{W}(t+\frac{y'-y}{v})},
\label{qlcWW}
\ee
\end{widetext}
where we explicitly indicate the dependence on the interaction parameter $K$.\\
In what follows we will first analyze the non-interacting case, which can be thought as a superposition of two independent integer QH systems subjected to opposite magnetic fields. Later we will address the case of interacting helical edge states.

\subsection{Non-interacting helical edge states}
\noindent
In the non-interacting case ($K=1$), one has $c^{(+)}_{K=1}=1$ and $c^{(-)}_{K=1}=0$, and Eq. \eqref{qlcWW} reduces to 
\beq
{\bf{\Gamma}}_{1}&=&\int d\vec{x}~d\vec{y} \frac{\Lambda_{x,y}\Lambda_{x',y'}^*}{(2\pi a)^2}e^{ik_F(y'-y+x'-x)}\nonumber \\
&\times&\int dt~e^{ieV t}e^{\mathcal{W}(t-\frac{x'-x}{v})+\mathcal{W}(t+\frac{y'-y}{v})}
\label{qlcWW2}
\eeq
where we introduced the short hand notation $d\vec{x}\equiv dx \cdot dx'$, $d\vec{y}\equiv dy \cdot dy'$.
In terms of the new variables \cite{Chevallier10} $\tau=t-\frac{y-y'-x+x'}{2v}$ and $z=\frac{y-y'+x-x'}{2}$ one has  
\begin{eqnarray}
{\bf{\Gamma}}_{1}&=&\int d\vec{x}~d\vec{y}~ \frac{\Lambda_{x,y}\Lambda_{x',y'}^*}{(2\pi a)^2}e^{i\left [k_{+}\left (x'-x\right )+k_{-}\left (y'-y\right ) \right ]} \nonumber \\
&\times&\int d\tau~ e^{ieV\tau}e^{\left [\mathcal{W}(\tau-\frac{z}{v})+\mathcal{W}(\tau+\frac{z}{v})\right ]}, \label{qlctau}
\end{eqnarray}
with $k_{\pm}= k_F\pm eV/2v$.
This can be further expressed as 
\be
{\bf{\Gamma}}_{1}=\int d\vec{x}~d\vec{y}~ \frac{\Lambda_{x,y}\Lambda_{x',y'}^*}{(2\pi a)^2}e^{i\left [ k_{+}(x'-x)+k_{-}(y'-y) \right ]}\tilde{F}_1(z,eV)
\label{rhs}
\ee
where
\begin{equation}
\tilde{F}_g(z,\omega)=\int d\tau~ e^{i\omega\tau}P_g\left (\tau-\frac{z}{v}\right )P_g\left (\tau+\frac{z}{v}\right )
\label{Pgz}
\end{equation}
and $P_g(t)=e^{g\mathcal{W}(t)}$ (cf. Eq. (\ref{W_exact})).\\
The separability assumption in Eq. (\ref{separabile}) allows to factorize the tunneling amplitude as
\begin{eqnarray}
&{\bf{\Gamma}}_{1}&=4\frac{\left |\Lambda_0\right |^2}{(2\pi a)^2}\int d\vec{y}~ \cos \left [\frac{eV}{v}\left (y'-y \right)\right ]f_c(|2y|)f_c(|2y'|) \nonumber \\
\label{rate4}
&\times& \int d\vec{x} \cos \left [ 2k_F\left (x'-x\right ) \right ]f_l(|2x|)f_l(|2x'|)\tilde{F}_1(x'-x,eV). \nonumber\\
\label{rate3b}
\end{eqnarray}
To better characterize the effects of the extended  contact geometry it is useful to represent ${\bf{\Gamma}}_{1}$ in terms of the point contact rate ${\bf{\Gamma}}^{(point)}_{1}$ as
\begin{equation}\label{Rate1}
{\bf{\Gamma}}_{1}=\lambda_{1}\times {\bf{\Gamma}}^{(point)}_{1}.
\end{equation}
This can be done regardless of the form of the tunneling amplitude but, as we will see, the separability assumption of Eq. (\ref{separabile}) allows to give a closed form for the modulating function.
From Eq. \eqref{current_detailed} and Eq. \eqref{Rate1} follows that
\begin{equation}\label{I_point}
\langle I_{\mathrm{BS}} \rangle=\lambda_1\times\langle I^{(point)}_{\mathrm{BS}} \rangle.
\end{equation}
For any interaction $K$, the point-like current is given by \cite{Strom09}
\begin{equation}
\langle I^{(point)}_{\mathrm{BS}}\rangle=2e(1-e^{-\beta eV}) \frac{\left |\Lambda_0\right |^2}{(2\pi a)^2}\tilde{P}_{2d_K}(eV)
\end{equation}
with $d_K\equiv c^{(+)}_K+c^{(-)}_K=\frac{1}{2}\left (K+\frac{1}{K}\right )$ so that $d_K=1$ in the non-interacting case.
The function 
\begin{equation}
\tilde{P}_g(\omega)=\int dt~e^{i\omega t}P_g(t)
\end{equation}
has the following form \cite{Overbosch09} for energies lower than the bandwidth $\omega_{c}$
\be
\tilde{P}_g(E)=\left\{
\ba \frac{2 \pi}{\Gamma(g)\omega_{c}}\left(\frac{E}{\omega_{c}}\right)^{g-1} \theta{(E)} \ \left (T=0\right )\\
\left (\frac{2\pi}{\beta\omega_c}\right )^{g-1}\frac{e^{\frac{\beta E}{2}}}{\omega_c} \mathcal{B}\left [\frac{g}{2}-i\frac{\beta E}{2\pi},\frac{g}{2}+i\frac{\beta E}{2\pi}\right ] \ \left (T\neq 0\right )
\ea\right.
\label{P_zero}
\ee
with $\theta(x)$ the Heaviside step function and $\mathcal{B}\left[x,y\right ]$ the Euler Beta function.\\
The modulating function $\lambda_{1}$ in Eq. \eqref{Rate1} represents the influence of the extended region and is given by
\begin{eqnarray}
\lambda_{1}&=&4\int d\vec{y}~ \cos \left [\frac{eV}{v}\left (y'-y \right)\right ]f_c(|2y|)f_c(|2y'|) \nonumber \\
\label{rate4b}
&\times& \int d\vec{x}~ \cos \left [ 2k_F\left (x'-x\right ) \right ]f_l(|2x|)f_l(|2x'|)\nonumber \\
&\times&\frac{\tilde{F}_1(x'-x,eV)}{\tilde{P}_2(eV)}. \nonumber \\
\label{coeff}
\end{eqnarray}
It can be written as a product of crossed and lateral contribution $\lambda_{1}=\lambda_{1}^{c}\lambda_{1}^{l}$, with
\begin{eqnarray}
\lambda^{c}_{1}&=&2\int d\vec{y}~ \cos \left [\frac{eV}{v}\left (y'-y \right )\right ]f_c(|2y|)f_c(|2y'|) \label{ampc} \\
\lambda^{l}_{1}&=&2\int d\vec{x}~ \cos \left [ 2k_F\left (x'-x\right ) \right ]f_l(|2x|)f_l(|2x'|)\nonumber \\
&\times &\frac{\tilde{F}_1(x'-x,eV)}{\tilde{P}_2(eV)}.
\label{ampl}
\end{eqnarray}
Notice that, while $\lambda^{c}_{1}$ depends on the crossed contribution $f_{c}$ only, $\lambda^{l}_{1}$ contains also the electronic Green's functions through $\tilde{F}_{1}$.\\
In order to perform an analysis of the extended contact, we consider a separable gaussian form \cite{Chevallier10, Overbosch09}
\begin{equation}
\Lambda_{x,y}=\frac{\Lambda_0}{2\pi\xi_c\xi_l}e^{-\frac{(x-y)^2}{4\xi_c^2}}e^{-\frac{(x+y)^2}{4\xi_l^2}}.
\label{gauss}
\end{equation}
The parameter $\xi_{l}$ is related to the extension of the contact, while $\xi_c$ allows to take into account non perfectly vertical events. In this sense a realistic assumption for modeling an extended contact is $\xi_c\ll\xi_l$. Note that in the limits $\xi_{c,l}\to 0$ we recover the point-like tunneling amplitude $\Lambda_{x,y}\to\Lambda_0\delta(x)\delta(y)$, so that $\langle I_{BS}\rangle\to\langle I^{(point)}_{BS}\rangle$.\newline
By replacing the gaussian expression into Eqs. (\ref{ampc})-(\ref{ampl}) one obtains
\beq
\lambda^{c}_{1}&=&e^{-\frac{1}{2}\left (\frac{\xi_c eV}{v} \right )^2} \label{ampc2} \\
\lambda^{l}_{1}&=&\frac{1}{\sqrt{2\pi}}\int dx e^{-\frac{x^2}{2}}\cos\left (2k_F\xi_lx\right )\frac{\tilde{F}_1(\xi_l x,eV)}{\tilde{P}_{2}(eV)}.
\label{ampl2}
\eeq
By exploiting the convolution properties
\be
\tilde{F}_g(z,\omega)=\frac{1}{2\pi}\int dE~ e^{i\frac{2z}{v}E}\tilde{P}_g\left (\frac{\omega}{2}+E\right )\tilde{P}_g\left (\frac{\omega}{2}-E\right ),
\label{Pgtil}
\ee
the tunneling amplitude can be written in the form
\beq
\lambda_{1}&=&e^{-\frac{1}{2}\left (\xi_c\frac{eV}{v}\right )^2-2(k_F\xi_l)^2}\int \frac{dE}{2\pi}e^{-2\left ( \xi_l\frac{E}{v} \right )^2}\cosh \left ( 4k_F\xi_l^2\frac{E}{v} \right )\nonumber\\
&\times&
\frac{\tilde{P}_{1}\left(\frac{eV}{2}+E\right) \tilde{P}_{1}\left(\frac{eV}{2}-E\right) }{\tilde{P}_{2}(eV)}. \label{rate_simplified}
\eeq
This result is valid also at finite temperature and extends what done in Ref. \onlinecite{Chevallier10} for the QHE at $T=0$. Note that the crossed contribution to the modulating function comes into play only at high bias voltage. For an extended contact with length $\sim (0.1 \div 1)\mu$m, one has $\xi_l\sim (0.1 \div 1) \mu$m and $\xi_c\ll \xi_l$, e.g. $\xi_c\sim 10$ nm. With this assumption the crossed contribution is crucial only for relatively high bias $\gtrsim 0.1$ V, not considered here. This fact allows to choose $\lambda^{c}_{1}\approx 1$ and to focus only on the lateral contribution which, as we will see in the following, shows strong modifications with respect to the point-like case also at low bias.

\subsection{Interacting helical edge states}

Starting from the general expression in Eq. (\ref{qlcWW}) and proceeding as in the previous section, one can express the interacting modulating function as ($K\neq 1$)
\begin{widetext}
\be
\lambda_{K}=\int \frac{ dE_{1} d E_{2} d E_{3}}{\left( 2 \pi\right)^{3}}e^{-\frac{1}{2} \left[ \frac{\xi_{c}}{v}\left(eV-2 E_{2}-2E_{3}\right)\right]^{2}-\frac{1}{2} \left[ \frac{\xi_{l}}{v}\left(eV-2 E_{1}-2E_{2}- 2k_{F} v\right)\right]^{2}}\frac{\tilde{P}_{c^{(+)}_K}(E_{1})\tilde{P}_{c^{(-)}_K}(E_{2})\tilde{P}_{c^{(-)}_K}(E_{3})\tilde{P}_{c^{(+)}_K}(eV-\underset{i=1,2,3}{\sum} E_i)}{\tilde{P}_{2 d_K}(eV)}
\label{lambda_general}
.\ee
\end{widetext}
Due to the natural constraints imposed by the functional form of $\tilde{P}(E)$ in Eq. (\ref{P_zero}) it is possible to neglect the crossed contribution, present in the first gaussian term, as far as $eV, k_{B}T\ll v/\xi_c$. Under this condition and noting that
\begin{equation}
\int_{-\infty}^{\infty}\frac{dE}{2\pi}\tilde{P}_{g_1}(E)\tilde{P}_{g_2}(\omega-E) =\tilde{P}_{g_1+g_2}(\omega),
\label{pezzo}
\end{equation}
Eq. (\ref{lambda_general}) becomes
\beq
\lambda_{K}&=&e^{-2\alpha_l^2}\int \frac{dE}{2\pi}e^{-2\left ( K\alpha_l\frac{E}{\epsilon_F} \right )^2}\cosh \left ( 4K\alpha_l^2\frac{E}{\epsilon_F} \right ) \nonumber\\
&\times&\frac{\tilde{P}_{d_K}\left(\frac{eV}{2}+E\right) \tilde{P}_{d_K}\left(\frac{eV}{2}-E\right) }{\tilde{P}_{2d_K}(eV)}. \label{rateverticale}
\eeq
Here, we introduced the Fermi energy $\epsilon_F=k_Fv_F$ and the dimensionless parameter $\alpha_l = k_F\xi_l$.
The modulating function thus depends on the length of the contact $\xi_l$ and on the Fermi momentum only through their product.
By inserting Eq. \eqref{P_zero} in Eq. \eqref{rateverticale} one has
\beq
\lambda_{K}&=&\frac{\Gamma (2 d_K) e^{-2\alpha_l^2}}{8 \pi^2\Gamma^2(d_K)}\!\!\int \!\!dx e^{-\frac{1}{2}(K\alpha_l\frac{k_BT}{\epsilon_F}x)^2}\!\!\cosh\left (2K\alpha_l^2\frac{k_BT}{\epsilon_F}x\right )\nonumber \\
&\times& \mathcal {B}\left[ \gamma_{+,+}(x), \gamma_{+,-}(x)\right]\mathcal {B}\left[ \gamma_{-,+}(x), \gamma_{-,-}(x)\right]
\label{ratefinale}
\eeq
with ($\eta, \eta '=\pm$) 
\be
\gamma_{\eta, \eta'}(x) = \frac{d_{K}}{2}+\eta\frac{i}{4\pi} \left(\frac{eV}{k_{B}T} +\eta' x\right).
\ee
To conclude we observe that also in the interacting case the backscattering current can be written as
\be\label{Ilong}
\langle I_{\mathrm{BS}}(V, T) \rangle= \lambda_{K}(V, T)\times \langle I^{(point)}_{\mathrm{BS}}(V, T) \rangle
\ee
with $\langle I^{(point)}_{\mathrm{BS}}(V, T) \rangle$ given in Eq. \eqref{I_point} and where we explicitly reintroduced the dependence on bias and temperature.
Note that for $\alpha_l=0$, Eq. \eqref{ratefinale} reduces to $\lambda_K=1$, and the point-like tunneling case is recovered.

\section{Results} \label{results}

\begin{figure}[!ht]
\centering
\includegraphics[scale=0.52]{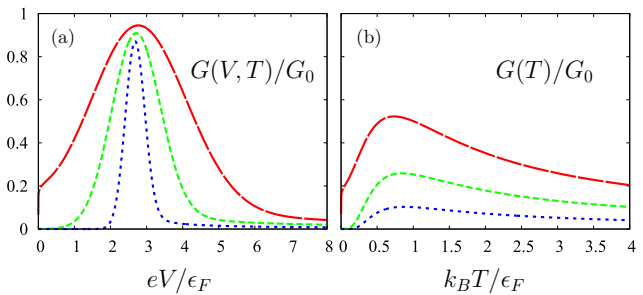}
\caption{(Color online) (a) Differential conductance as a function of bias $V$ (in units of $\epsilon_F/e$) at low temperature ($k_BT=10^{-2}\epsilon_F$) and (b) linear conductance as a function of the temperature $T$ (in units of $\epsilon_F/k_B$), for different lengths of the contact: 
$\alpha_l=1$ (long dashed red), $2$ (dashed green), $5$ (short dashed blue).
Units of the conductance: $G_0=\frac{2 e^2}{\epsilon_F^2}\frac{\left | \Lambda_0\right |^2}{(2\pi a)^2}\left (k_F a\right )^{2d_K}$.
Other parameters: $K=0.75$.}\label{conduttanze}
\end{figure}

\begin{figure}[!ht]
\centering
\includegraphics[scale=0.40]{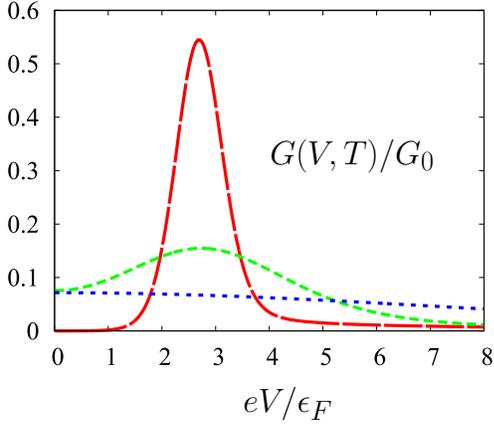}
\caption{(Color online) Differential conductance as a function of bias $V$ (in units of $\epsilon_F/e$) for different temperatures (in units of $\epsilon_F/k_{B}$): $T=0.1$ (long dashed red), $T=0.5$ (dashed green), $T=2$ (short dashed blue).
Units of $G_0$ as in Fig. \ref{conduttanze}.
Other parameters: $\alpha_l=5$; $K=0.75$.}\label{G_T}
\end{figure}

Since the modulating function depends on bias and temperature, it will influence the behavior of transport properties with respect to the point-like tunneling case.
It is then useful to investigate it in details.
\begin{figure}[!ht]
\centering
\includegraphics[scale=0.52]{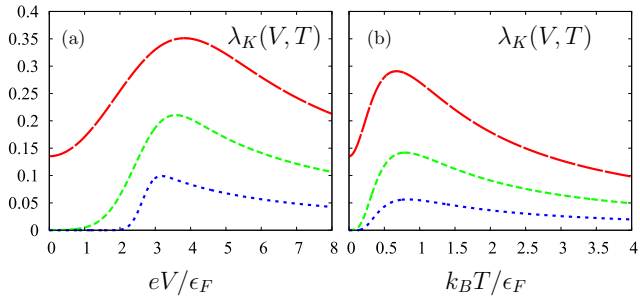}
\caption{(Color online) Modulating function as a function of (a) bias $V$ (in units of $\epsilon_F/e$) at low temperature ($k_BT=10^{-2}\epsilon_F$) and (b) temperature $T$ (in units of $\epsilon_F/k_B$) at low bias ($eV=10^{-2}\epsilon_F$), for different lengths of the contact: 
$\alpha_l=1$ (long dashed red), $2$ (dashed green), $5$ (short dashed blue). Note that the behavior at low temperature in (a) is indistinguishable from the $T=0$ case. This comment holds as well for panel (b) between low $V$ and $V=0$. Other parameters: $K=0.75$.}\label{lambda}
\end{figure}
Fig. \ref{lambda} shows $\lambda_K$ as a function of voltages (a) or temperatures (b).
Fig. \ref{lambda}(a) presents a maximum at $V\approx \bar{V}\equiv 2\epsilon_F/eK$, becoming more and more pronounced by increasing $\alpha_l$, that is the length of the contact. In the limit $\alpha_l\to 0$ it is washed out and $\lambda_K(V, T)\to 1$.
As already noted for QHE \cite{Overbosch09}, this maximum is determined by the two phases that control tunneling, one set by the Fermi momentum ($2k_Fx$) and the other by the voltage drop ($eVt$). The peak occurs when the two phases are equal: $e\bar{V}=2k_F x/t=2k_Fv=2\epsilon_F/K$.\\
A maximum is present also in Fig. \ref{lambda}(b), but it originates from a dephasing mechanism, induced by finite temperature, similar to what was found in interferometric geometries with two or several QPCs, both in QH \cite{Chamon97} and in QSH systems \cite{Virtanen11}, where the dephasing was depending on the distance among the QPCs.
The extended contact geometry can be seen indeed as an infinite series of QPCs with different tunneling amplitudes, with infinitesimal distance $dx$ between them, and the backscattering current is now given by integrating over the contact region.
For all interaction strengths $0.5 < K < 1$ we find the maximum at a position $\bar{T}$ of the order of $\epsilon_F/k_B$, vanishing as $\alpha_l\to 0$, reproducing in this case the point-like regime with $\lambda_K(V,T)\to 1$.\\
Note that for vanishing bias and temperature the modulating function is exponentially suppressed by the length of the contact, namely $\lambda_K(V=0,T=0)=e^{-2\alpha_l^2}$.\\
We can also study the asymptotic behavior of $\lambda_K$ at low bias or low temperatures. Introducing the energy scales $eV_{\alpha_l}=\epsilon_F/(K\alpha_l)$ and $k_BT_{\alpha_l}=\epsilon_F/(K \alpha_l)$ one finds
\begin{equation}\label{asympt_V}
\lambda_K(V,T\ll eV/k_B)\sim\left\{
\begin{array}{cc}
\mathrm{constant} & V\ll V_{\alpha_l} \\
V^{-1} & V-\bar{V}\gg V_{\alpha_l}
\end{array}
\right.
\end{equation}
and
\begin{equation}\label{asympt_T}
\lambda_K(V\ll k_BT/e,T)\sim\left\{
\begin{array}{cc}
\mathrm{constant} & T\ll T_{\alpha_l}  \\
T^{-1} & T-\bar{T}\gg T_{\alpha_l}
\end{array}
\right. .
\end{equation}
Fig. \ref{conduttanze} shows the differential conductance $G(V,T)=d\langle I_{\mathrm{BS}}(V, T)\rangle/dV$ as a function of bias (a) and the linear conductance $G(T)=G(V=0,T)$ as a function of temperature (b).
They both show a peaked structure, in contrast to the point-like case, reminiscent of the form of $\lambda_K$ (see Fig. \ref{lambda}).\\
More quantitatively, focusing on a given length, we can study the dependence on interactions.
Fig. \ref{G_T=0,K} shows the differential conductance as a function of bias, varying the electron interaction.
\begin{figure}[!ht]
\centering
\includegraphics[scale=0.40]{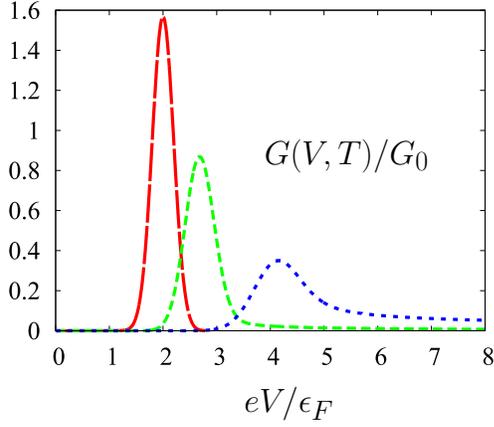}
\caption{(Color online) Differential conductance as a function of bias $V$ (in units of $\epsilon_F/e$) for different interaction strengths: $K=1$ (long dashed red), $0.75$ (dashed green), $0.5$ (short dashed blue).
Note that the conductance is plotted in unity of $G_0$ as in Fig. \ref{conduttanze}, which depends on $K$ and thus not allow for a direct comparison on the size between the different curves.
Other parameters: $\alpha_l=5$; $k_BT=10^{-2}\epsilon_F$.}\label{G_T=0,K}
\end{figure}
The conductance shows a peak at $V\approx \bar{V}$, which depends on the velocity of the excitations ($\bar{V}=2 k_Fv/e$).
Thanks to this behavior, we argue that an extended contact geometry could be fruitful to extract information about the velocity of the excitation modes along the edges, by experimentally measuring the peak of the conductance, varying the Fermi energy and the bias voltage \cite{Konig07, Roth09}.
Furthermore, it must be stressed that in presence of an ordinary Luttinger liquid we should expect two different peaks, as a consequence of the spin-charge separation, which leads to two different propagation velocities, one for the charge modes and one for the spin modes \cite{Giamarchi03, Miranda03, Auslander05, Cavaliere04}.
The single-peak structure of Fig. \ref{G_T=0,K}, instead, provides evidence of the close connection between spin and charge typical of the helical edge states of QSHE, where these degrees of freedom are locked each other and propagate with the same velocity.\\
We remark that, as expected, the peak in the differential conductance is reduced by increasing temperature and finally washed out for temperatures $2\pi k_BT\sim e\bar{V}$, as shown in Fig. \ref{G_T}.
\begin{figure}[!ht]
\centering
\includegraphics[scale=0.50]{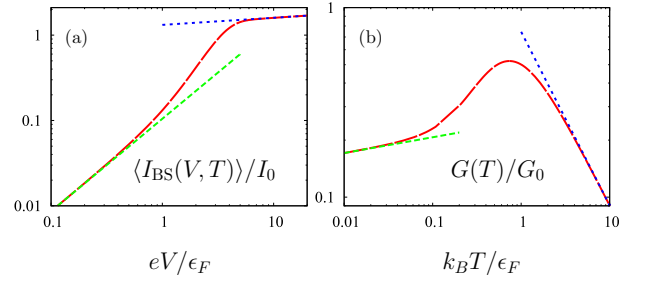}
\caption{(Color online) Log-Log plot (a) of the backscattering current (in units of $I_0\equiv (\epsilon_{F}/e) G_{0}$) as a function of the bias voltage $V$ (in units of $\epsilon_F/e$) at low temperature ($k_{B}T=10^{-2} \epsilon_F$) (long dashed red curve) and (b) of the linear conductance (in units of $G_0$) as a function of temperature (in units of $\epsilon_F/k_B$) (long dashed red curve).
Other parameters: $\alpha_{l}=1$, $K=0.75$.
Straight lines represent the asymptotic power-law behavior with exponent (a) $2d_{K}-1=13/12$ (dashed green line) and $2d_{K}-2=1/12$ (short dashed blue line) and (b) $2d_{K}-2=1/12$ (dashed green line) and $2d_{K}-3=-11/12$ (short dashed blue line). }\label{power_law}
\end{figure} \\
Information about velocity are not the only ones that can be extracted by means of this setup. Theoretical works concerning point-like tunneling predict power-law behaviors for current \cite{Strom09, Schmidt11}
\begin{equation}
\langle I^{(point)}_{\mathrm{BS}}(V, T\ll eV/k_B)\rangle\sim V^{2d_K-1}
\end{equation}
\begin{equation}
G^{(point)}(T)\sim T^{2 d_K-2}.
\end{equation}
Despite these trends are here no longer valid, they still survive at bias or temperatures lower enough, namely for $V\ll V_{\alpha_l}$ or $T\ll T_{\alpha_l}$ respectively, as shown in Fig. \ref{power_law}. Interestingly, by increasing energies, new power-law behaviors are recovered, however, with different exponents
\begin{equation}
\langle I_{\mathrm{BS}}(V, T\ll eV/k_B)\rangle\sim
V^{2d_K-2} \qquad (V-\bar{V}\gg V_{\alpha_l})
\end{equation}
and
\begin{equation}
G(T)\sim T^{2d_K-3} \qquad (T-\bar{T}\gg T_{\alpha_l}).
\end{equation}
This is a consequence of the asymptotic behavior of the modulating function (cf. Eqs. \eqref{asympt_V}-\eqref{asympt_T}).\\
It is worth noting that the effective visibility of these high energy power-laws crucially depends on the Fermi energy $\epsilon_{F}$ of the system, that can be easily tuned  experimentally by means of an external gate \cite{Konig07}, and the natural cut-off energy $\omega_c$ of the theory. The latter can be reasonably identified as the energy at which additional bulk effects have to be taken into account, thus the presented helical Luttinger liquid picture holds for energies lower than $\omega_c$.

\section{Conclusions} \label{conclusion}
We proposed a model for an extended tunneling through contact region in QSH system. We demonstrated that it is possible to take into account the extended nature of the contact through a modulating function, which renormalizes the transport properties of the point-like case.\\
We showed that, due to the extended nature of the contact and for low enough temperatures, the differential conductance shows a pronounced peak that can be used to extract information about the propagation velocity of the excitations along the edge. The presence of a unique peak is a signature of the helical nature of the edge states in QSHE.\\
We analyzed the backscattering current in the low temperature regime and the linear conductance, showing that the power-law behaviors predicted in the point-like case survive at progressively lower energies by increasing the length of the contact. Remarkably enough, new power-laws emerge also at higher energies, but with different exponents.

\section*{Acknowledgements}

We thank A. Braggio, M. Carrega, and T. Martin for useful discussions. The support of CNR STM 2010 program, EU-FP7 via Grant No. ITN-2008-234970 NANOCTM and CNR-SPIN via Seed Project PGESE001 is acknowledged.

\end{document}